\documentclass[aps,a4paper,amssymb,amsmath,10pt,twocolumn,superscriptaddress]{revtex4}
\usepackage[cp1251]{inputenc}
\usepackage[dvips]{graphicx,color}
\usepackage{psfrag}
\usepackage[english]{babel}
\definecolor{darkgreen}{rgb}{0,0.7,0}
\definecolor{darkblue}{rgb}{0,0,0.7}
\definecolor{darkred}{rgb}{0.7,0,0}
\definecolor{brightred}{rgb}{1,0.1,0.1}
\definecolor{greenblue}{rgb}{0,0.45,0.4}
\usepackage[dvips,unicode, colorlinks, citecolor=darkblue, linkcolor=darkred, urlcolor=blue]{hyperref}

\newcommand{\varint}[0]{\mathop{\int\!\!}\limits}
\newcommand{\sio}[0]{{\rm SiO$_2$ }}
\newcommand{\tao}[0]{{\rm Ta$_2$O$_5$ }}

\begin{document}
\title{Reducing Thermal Noise in Future Gravitational Wave Detectors by employing Khalili Etalons}
\author{Alexey G. Gurkovsky}
\affiliation{Faculty of Physics, Moscow State University, Moscow, 119991 Russia}
\author{Daniel Heinert}
\affiliation{Institut f\"ur Festk\"orperphysik, Friedrich-Schiller-Universit\"at Jena, D-07743 Jena, Germany}
\author{Stefan Hild}
\affiliation{SUPA, School of Physics and Astronomy, Institute for Gravitational Research, Glasgow University, Glasgow G12 8QQ, United Kingdom}
\author{Ronny Nawrodt}
\affiliation{Institut f\"ur Festk\"orperphysik, Friedrich-Schiller-Universit\"at Jena, D-07743 Jena, Germany}
\author{Kentaro Somiya}
\affiliation{Waseda Institute for Advanced Study, 1-6-1 Nishiwaseda, Shinjuku, Tokyo 169-8050, Japan}
\affiliation{Interactive Research Center of Science, Tokyo Institute of Technology, 2-12-1 Oh-okayama, Meguro, Tokyo 152-8551, Japan}
\author{Sergey P. Vyatchanin}
\affiliation{Faculty of Physics, Moscow State University, Moscow, 119991 Russia}
\author{Holger Wittel}
\affiliation{Max Planck Institute for Gravitational Physics (Albert Einstein Institute), D-30167 Hannover, Germany}
\date{\today}

\begin{abstract}
Reduction of thermal noise in dielectric mirror coatings is a key issue for the sensitivity improvement in second and third generation interferometric gravitational wave detectors. Replacing an end mirror of the interferometer by an anti-resonant cavity (a so-called Khalili cavity) has been proposed to realize the reduction of the overall thermal noise level. In this article we show that the use of a Khalili etalon, which requires less hardware than a Khalili cavity, yields still a significant reduction of thermal noise. We identify the optimum distribution of coating layers on the front and rear surfaces of the etalon and compare the total noise budget with a conventional mirror. In addition we briefly discuss advantages and disadvantages of the Khalili etalon compared with the Khalili cavity in terms of technical aspects, such as interferometric length control and thermal lensing.

\end{abstract}

\maketitle
\section{Introduction}
The sensitivities of second-generation (Advanced LIGO, Advanced VIRGO, GEO-HF, and LCGT) and third-generation (Einstein Telescope) interferometric gravitational wave detectors will be partly limited by thermal fluctuations in the mirrors \cite{2006_LIGO_status,2006_VIRGO_status,2006_GEO-600_status,2006_LCGT_status,2010_ET_status}.

The pioneering articles on this issue were dedicated to the investigation of the mirror {\em substrate} fluctuation: Brownian thermal noise \cite{1995_BrownCoat_Raab, 1998_FDT_Levin, 1998_ThermalNoise_Bondu} and thermo-elastic noise~\cite{1999_ShotNoise_BGV}. Fundamental thermal motion (Brownian motion) of material atoms or molecules causes Brownian noise. Fundamental thermodynamic fluctuations of temperature lead to thermo-elastic noise through the material's thermal expansion. Similarly thermo-refractive noise \cite{2000_TRNoise_BGV,2009_TRNoise_Levin} is caused by temperature fluctuations leading to fluctuations of the refractive index and therefore fluctuations of the optical path length inside the material. These results were obtained for the model of an infinite test mass, i.e. the mirror was considered to be an elastic layer with infinite width and finite thickness. All of these results were generalized for a finite-size mirror model \cite{1998_ThermalNoise_Bondu,2000_TENoise_Thorne}.
 
\begin{figure*}[htbp]
\begin{center}
\includegraphics[width=1\textwidth]{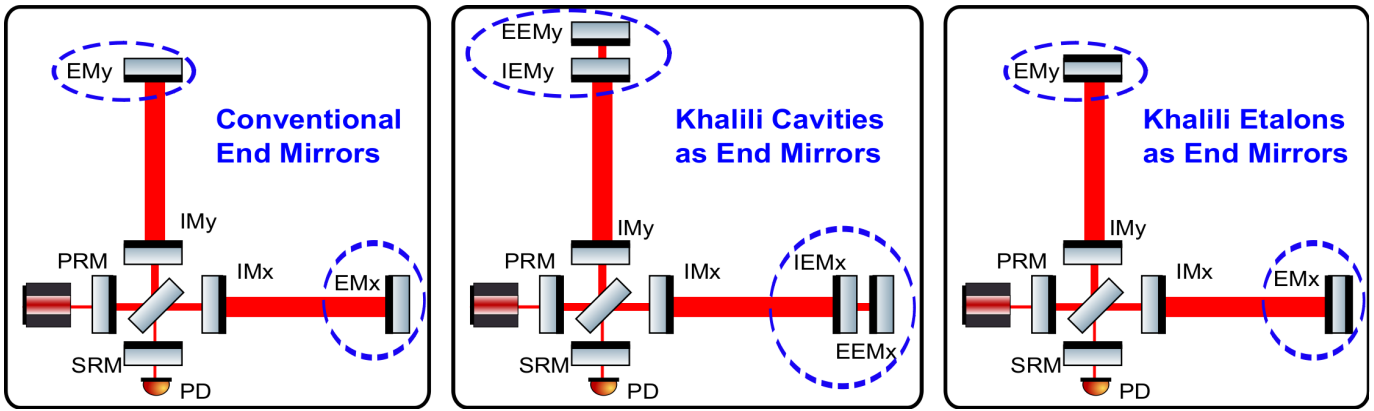}
\end{center}
\caption{Simplified schematic of an Advanced LIGO interferometer with conventional 
end mirrors, featuring all the coating layers on their front (left), with Khalili cavities as
end mirrors (center) and with Khalili etalons, featuring only a few coating layers on the front 
surface and the majority of the coating layers on the rear surface (right).}\label{fig:schematics}
\end{figure*}

Very soon the importance of mirror {\em coating} thermal noise was realized as its parameters may differ considerably from the mirror substrate parameters. Despite its low thickness, the very high loss angle of the mirror coating materials (usually \sio\ and Ta$_2$O$_5$)
 makes the coating Brownian noise the most significant one among all kinds of mirror thermal noise \cite{2002_ThermCoat_Harry,2006_ThermCoat_Harry,2009_ThermCoatCyl_Somiya}. The thermo-elastic noise of the coatings only has a small contribution in the total noise budget \cite{2003_CoatTDFluct_BV,2004_TDInhomog_Fejer}. Later, Kimble \cite{2008_BrownCompensate_Kimble} proposed the idea of thermal noise compensation which was explored carefully in \cite{2009_TermoOpt_Evans, 2008_TDCompenstae_Gorodetsky} for a particular case of thermo-elastic and thermo-refractive noise.

The coating Brownian noise is still one of the main contributions to the noise spectra of gravitational wave observatories \cite{2002_ThermCoat_Harry,2006_ThermCoat_Harry,2009_ThermCoatCyl_Somiya}. One of the most promising approaches aimed to decrease its level was offered by Khalili \cite{2005_DM_Khalili} who proposed to replace the end mirror in the interferometer arm with a short Fabry-P\'erot cavity tuned to anti-resonance (see center panel of Figure \ref{fig:schematics}). In practice most light is reflected from only a few first layers (farest from substrate) and all others (located closer to substrate) only reflect a small part of light. However, since the thermal fluctuations are proportional to the total thickness of the coating and the inner layers of coating are the main contribution to the phase fluctuations of the reflected light, the transmittance of each mirror can be higher to realize the same reflectivity of the system as a compound end mirror. The total thickness of coatings in such {\em Khalili cavity} mirrors is the same as the thickness of the conventional mirror, while the Brownian noise of the end mirror of a Khalili cavity is significantly reduced \cite{2005_DM_Khalili} because the thickness fluctuations of the second end mirror coating (EEM in Figure \ref{fig:schematics}) do not influence the fluctuations of the input mirror coating (IEM in Figure \ref{fig:schematics}). Moreover, using a rigidly controlled Khalili cavity allows a reduction of 
 coating Brownian noise \cite{2009_DM_Somiya}. 

One of the main problems in the Khalili cavity is to establish a low-noise control of the mirror positions (see detailed explanation in Sec.~\ref{ThermLens}). A potentially easier way is to use a {\em Khalili etalon} (KE) instead of a Khalili cavity (KC) or a simple conventional mirror (CM). The idea is to use a single mirror but to split the coating into two parts (see right hand panel of Figure (EEM in Figure \ref{fig:schematics}): {\em the front coating} (on the front substrate surface) features just a few layers and {\em the rear coating} (on the rear substrate surface) consists of the rest of the required coating layers. 

The purpose of this article is to develop an idea of the Khalili etalon \cite{2010_KE_Somiya}, to calculate the total mirror thermal noise arising in a KE and in a CM, and to compare them. We investigate the idea of using a KE in the Einstein Telescope (ET) and Advanced LIGO (aLIGO). In Sec.~\ref{Optim} we describe the mirror parameter optimization procedure, namely the optimal number of layer pairs in the front coating. In Sec.~\ref{TNinKE} we describe the details of the thermal noises arising in the KE and CM calculations. Section~\ref{ThermLens} is dedicated to the problem of thermal lensing which is much more important in a KE than in a CM. In Sec.~\ref{Conc} we discuss the obtained results and draw the conclusions. Finally, some calculation details are provided in the appendices \ref{AppA}-\ref{AppB}.

\section{Coating optimization}\label{Optim}
The main idea of using a KE is to reduce the mirror's total thermal noise without reducing its reflectivity. By {\em total thermal noise spectral density} we mean the sum of the Brownian, thermo-elastic and thermo-refractive noise spectral densities. Coating Brownian noise is caused mostly by the fluctuations of the entire coating thickness. It would then seem evident that Brownian noise be lower when the front coating contains less layers and hence the lowest noise be achieved for the coating totally displaced to the rear mirror surface. This is in principle true but at the same time some other noises, such as substrate thermo-refractive noise, rise dramatically causing the total noise level to rise also. Moreover, the less layers one puts onto the front coating, the higher will be the absorption in the substrate. So there has to be an optimum of how to best distribute the coating layers between the front and back surfaces in order to obtain minimal total thermal noise and not too much of absorption in substrate. The aim of this section is to find this optimum configuration.
\subsection{Thermal Noise calculation technique}\label{N1Opt}
The only way to find the optimal number of front coating layers, $N_1$, is to compare the thermal noise for every $N_1$. This requires the calculation of the different noise contributions as functions of the front coating layers number $N_1$. The most basic principles we used are: (i) the total number of \tao\ and \sio\ layers, $N$, is fixed, i.e. we used the coating structure planned for both ET and aLIGO and modified it to fit the double coating paradigm: 20 \tao\ and 18 \sio\ quarter-wave layers plus the substrate (it is considered as an ordinary but ``slightly'' thicker coating layer) and plus two caps consisting of a half-wavelength \sio\ layer (for the CM it would have been 20 \tao\ layers and 19 \sio\ layers plus one cap); (ii) a quarter-wavelength \tao\ layer and a quarter-wavelength \sio\ layer are alternately coated on the front or rear surface so that there are always an odd number of front coating layers ($N_1=$ 1, 3, 5 etc.) and also an odd number of layers of the rear coating $N_2=N-N_1$ (37, 35, 33 etc.). Please note that the substrate and caps are not included in these numbers; (iii) the number of layers of the front coating $N_1$ is the argument and the total thermal noise driven mirror displacement $S_\text{total}^\text{(KE)}$ is the function of it; (iv) we consider only Brownian, thermo-elastic and thermo-refractive noises (being the most significant contributions), and (v) we used the mirror of a finite-size cylinder, the model of which has been developed in Ref.~\cite{1998_ThermalNoise_Bondu,2000_TENoise_Thorne,2009_ThermCoatCyl_Somiya,2010_KE_Somiya}, and calculated all noises numerically using the fluctuation-dissipation theorem (FDT) \cite{1998_FDT_Levin,1951_Callen_Welton,1986_Landafshitz} as it is briefly described in Secs.~\ref{optBN}-\ref{optAbs}. The optimal number of front coating layers appeared to be $N_1=3$, i.e. $2$ \tao\ layers and $1$ \sio\ layer plus a cap in the front coating and $18$ \tao\ layers and $17$ \sio\ layers plus a cap in the rear coating. With the technical feasibility taken into account (see Sec.~\ref{optAbs}), however, it turns out that the system with $N_1=5$ layers on the front surface (i.e. $3$ \tao\ layers and $2$ \sio\ layers plus a cap on the front mirror surface and $17$ \tao\ layers and $16$ \sio\ layers plus a cap on the rear surface) will be better and we analyze the system with $N_1=5$ in detail. In this case the mirror thermal noise does not reach its minimum but it is only about $5$\,\% higher.
\subsubsection{Brownian noise}\label{optBN}

The total coating thermal noise of the etalon will be the sum of noise on the front surface and noise on the back surface:
\begin{eqnarray}
\delta x=\epsilon_1 \delta x_1+\epsilon_2\delta x_2\ .
\end{eqnarray}
Here $\delta x_1$ is the displacement of the front surface of the mirror and $\delta x_2$ the displacement of the boundary surface between the rear surface of the mirror substrate and the coating on it. Considering the KE as a Fabry-P\'erot cavity consisting of two mirrors with amplitude reflectivities $R_1$ (front coating) and $R_2$ (rear coating) tuned to anti-resonance, one can calculate the coefficients $\epsilon_1$ and $\epsilon_2$ (see details in Ref.~\cite{2010_KE_Somiya}):
\begin{eqnarray}
\epsilon_1&=&\frac{R_1\left[1+(1+n_s)R_1R_2+R_2^2\right]+R_2(1-n_s)}{(1+R_1R_2)^2}\ ,\nonumber\\
\label{e1e2}
\epsilon_2&=&\frac{n_sR_2(1-R_1^2)}{(1+R_1R_2)^2}\ .
\end{eqnarray}
Here $n_s$ is the substrate refractive index. Note that $R_1$ and $R_2$ are functions of the number of front and rear coating layers. In particular, we have the following formulas for $R_1$ and $R_2$ as functions of the number of the front coating layers $N_1$ ($N_2=N-N_1$ is the number of the rear coating layers):
\begin{align}
\label{R1R2}
R_1 &= \frac{1-\dfrac{n_s n_2^{N_1-1}}{n_1^{N_1+1}}}{1+\dfrac{n_s n_2^{N_1-1}}{n_1^{N_1+1}}},\quad
R_2 = \frac{1-\dfrac{n_s n_2^{N-N_1-1}}{n_1^{N-N_1+1}}}{1+\dfrac{n_s n_2^{N-N_1-1}}{n_1^{N-N_1+1}}},
\end{align}
where $n_1$ and $n_2$ are \tao\ and \sio\ coating layers refractive indices.

Hence, in order to calculate spectral density $S(\omega)$ of the displacement $\delta x$ caused by thermal noise using the FDT, one has to apply the forces
\begin{equation*}
 \epsilon_1F_0e^{i\omega t}\quad \text{and}\quad \epsilon_2F_0e^{i\omega t}
\end{equation*}
to the front and rear coatings correspondingly and to calculate the total dissipated power \cite{1998_FDT_Levin, 1951_Callen_Welton, 1986_Landafshitz}. 
For the calculation of the spectral density $S_{B}(\omega)$ of Brownian coating noise the dissipated power may be calculated through the elastic energy $U_\text{B}^{(k)}$ stored in each $k$-th layer (of the front or rear coating):
\begin{multline}
U_\text{B}^{(k)}=\pi\varint_{h_k}dz\int_0^R\left(E_{rr}^{(k)}T_{rr}^{(k)}+E_{\phi\phi}^{(k)}T_{\phi\phi}^{(k)}+\right.\\
\left.+E_{zz}^{(k)}T_{zz}^{(k)}+E_{rz}^{(k)}T_{rz}^{(k)}\right)\,dr\, ,
\end{multline}
where $E_{ij}$ and $T_{ij}$ are the strain and stress tensor components (only the non-zero components are shown in the formula above), $R$ is the mirror radius and $h_k$ is the thickness of the $k$-th layer. The components $E_{ij}$ and $T_{ij}$ are calculated as it is described in detail in~\cite{2010_KE_Somiya}. Then the Brownian noise spectral density may be evaluated as follows:
\begin{equation}
S_\text{B}=\frac{8k_BT}{\omega}\sum_{k=0}^{N+2}U_B^{(k)}\phi_k
\end{equation}
where $k_B$ is Boltzmann's constant, $T$ is the absolute temperature and $\phi_k$ is the loss angle describing structural losses in the $k$-th layer. The sum is taken over all layers, i.e. $N=38$ is the number of layers without the substrate and the caps. The total number of summands is therefore $N+3=41$; $N=38$ layers in the front and rear coatings, plus $2$ layer-caps and $1$ layer-substrate. So the index $k=0$ refers to the front coating cap, the indices $k=1\div{N_1}$ refer to the front coating quarter wavelength (QWL) layers, the index $k={N_1+1}$ refers to the substrate, the indices $k={N_1+2}\div{N+1}$ refer to the rear coating QWL layers and the index $k={N+2}$ refers to the rear coating cap. There are $N+2=40$ summands; the ones with $k\neq N_1$ are to be considered for coating Brownian noise and the one with index $k={N_1+1}$ is to be considered for substrate Brownian noise:
\begin{subequations}
\begin{align}
S_\text{B}^\text{(coat)}&=\frac{8k_BT}{\omega}\left(\sum_{k=0}^{N_1}U_B^{(k)}\phi_k+\sum_{k=N_1+2}^{N+2}U_B^{(k)}\phi_k\right),\\
S_\text{B}^\text{(sub)}&=\frac{8k_BT}{\omega}U_B^{(N_1+1)}\phi_{N_1+1}.
\end{align}
\end{subequations}
Here $\phi_{N_1+1}=\phi_s$ is the loss angle of the substrate, while $\phi_1$ and $\phi_2$ represent the loss angels of the \tao\ and \sio\ layers, respectively. The values are presented in Table~\ref{paramPhys}.
\subsubsection{Thermo-elastic noise}\label{optTEN}
The thermo-elastic (TE) noise calculations for the substrate and for the coating is similar. In order to calculate the dissipated power one should calculate the diagonal components $E_{jj}$ of the strain tensor for each layer (including the substrate and the caps) and take the trace:
\begin{equation*}
\theta^{(k)}=E_{rr}^{(k)}+E_{\phi\phi}^{(k)}+E_{zz}^{(k)}.
\end{equation*}
Then one may find the power dissipated through the TE mechanism \cite{2000_TENoise_Thorne,1986_Landafshitz}:
\begin{equation}
W_\text{TE}^{(k)}=2\pi\kappa_k T \left(\frac{Y_k\alpha_k}{(1-2\nu_k)C_k\rho_k}\right)^2\varint_{h_k}dz
 \int_0^R\left[\vec\nabla\theta^{(k)}\right]^2r\,dr\, ,
\end{equation}
where $\kappa_k$ is the thermal conductivity, $C_k$ is the thermal capacity per unit volume, $Y_k$ is the Young's modulus, $\nu_k$ is the Poisson's ration, $\alpha_k$ is the thermal expansion coefficient, $\rho_k$ is the density, and the index $k$ denotes number of the layer. Therefore, the TE noise spectral density will simply be:
\begin{equation}
S_\text{TE}=\frac{8k_BT}{\omega^2}\sum_{k=0}^{N+2}W_{TE}^{(k)}
\end{equation}

Similar to the Brownian noise calculations, the summands with the indices $k\neq {N_1+1}$ are relevant for the coating TE noise while the one with the index $k={N_1+1}$ needs to be considered for the TE noise of the substrate:
\begin{subequations}
\begin{align}
S_\text{TE}^\text{(coat)}&=\frac{8k_BT}{\omega^2}\left(\sum_{k=0}^{N_1}W_{TE}^{(k)}+\sum_{k=N_1+2}^{N+2}\!\!\!W_{TE}^{(k)}\right),\\
S_\text{TE}^\text{(sub)}&=\frac{8k_BT}{\omega^2}W_{TE}^{(N_1+1)}.
\end{align}
\end{subequations}
\subsubsection{Thermo-refractive noise}\label{optTRN}

TR noise originates from thermodynamic fluctuations of the temperature $\delta T$ in the substrate, producing phase fluctuations of the reflected wave phase via the temperature dependence of the substrate's refraction index $n_s$. Likewise, the phase fluctuations may be recalculated into effective fluctuations of mirror surface displacement $\delta x=-\epsilon_2\beta_sh\delta T$ where the coefficient $\epsilon_2$ introduced in (\ref{e1e2}), characterizes the light amplitude circulating inside the substrate and $\beta_s= dn_s/dT$ is the thermo-optic coefficient of the substrate.

We calculate the thermo-refractive (TR) noise in the substrate using the model of an infinitely large plane in the transverse directions with thickness of $h$. The spectral density of the temperature fluctuations in this model is shown in \cite{2004_CornerRefl-TR_BV} see Eq.~(E8):
\begin{equation*}
S_\text{T}=\frac{16k_BT^2\kappa_s}{\pi\rho_s^2C_s^2w^4\omega^2h}
\end{equation*}
where $w$ is the radius of the light spot (intensity decreases with distance $r$ from center as $\sim e^{-2r^2/w^2}$) and the parameters with subscript $s$ refer to the substrate. The TR noise spectral density for the substrate (recalculated to displacement) becomes
\begin{equation}
S_\text{TR}^\text{(sub)}=\epsilon_2^2\beta_s^2h^2S_T=\epsilon_2^2\frac{16k_BT^2\beta_s^2\kappa_sh}{\pi\rho_s^2C_s^2w^4\omega^2}
\end{equation}

Benthem and Levin have pointed out that some corrections should be applied to this formula. This corrections are based on the account of the fact that light inside the arm froms the standing wave and not a traveling wave. We can rewrite Eq.~(2) of Ref.~\cite{2009_TRNoise_Levin} in a simpler form with only the normal incidence and the circular beam being considered:
\begin{equation}
S_\text{TR}^\text{(sub)}=\epsilon_2^2\frac{16k_BT^2\beta_s^2\kappa_sh}{\pi\rho_s^2C_s^2w^4\omega^2}\left(1+\frac{k^2w^2}{2(1+(2k\sqrt{\kappa_s/C_s\rho_s\omega})^4)}\right)
\end{equation}

In addition we have to consider the TR noise present in the coatings \cite{2000_TRNoise_BGV}. For its estimate we use the following formula
\begin{align}
S_\text{TR}^\text{(coat)} &=\frac{2\sqrt{2}k_BT^2\beta_\text{eff}^2\Lambda^2}{\pi
 \sqrt{\kappa_s\rho_sC_s}w^2\sqrt{\omega}},\\
\beta_\text{eff} &=\frac{1}{4}\frac{\beta_1n_2^2+\beta_2n_1^2}{n_1^2-n_2^2}
\end{align}
where $\Lambda$ is the wavelength of light in vacuum, $\beta_\text{eff}$ is the averaged thermo-optic coefficient of the entire coating, and $\beta_1=dn_1/dT$ and $\beta_2=dn_2/dT$ are the thermo-optic coefficients of \tao\ and \sio\ layers, respectively. This formula is based on the assumption that only the first few layers contribute considerably to the thermo-refractive loss mechanism. It is obtained for a mirror with an infinite radial dimension and a finite height. This model is valid with good accuracy for CM. However, for KE we use the same formula as an order-of-magnitude estimation.


\subsection{Optimization results}\label{optRes}

In this subsection we present the results of our optimization process. 
Using the proposed parameters we obtained numerical estimates of all noise sources discussed above for ET and aLIGO. 
All geometrical design parameters for these interferometers are presented in Table~\ref{paramGeom}. 
The physical constants and material parameters are summarized in Table~\ref{paramPhys}.

\begin{figure}[htbp]
\begin{center}
\psfrag{Einstein Telescope and Advanced LIGO to CM gain}[cc][cc]{\small }
\psfrag{KE to CM total noise gain}[cc][cc]{\small gain $G$}
\psfrag{Number of front coating layers}[cc][cc]{\small Number of front coating layers $N_1$}
\includegraphics[width=0.4\textwidth]{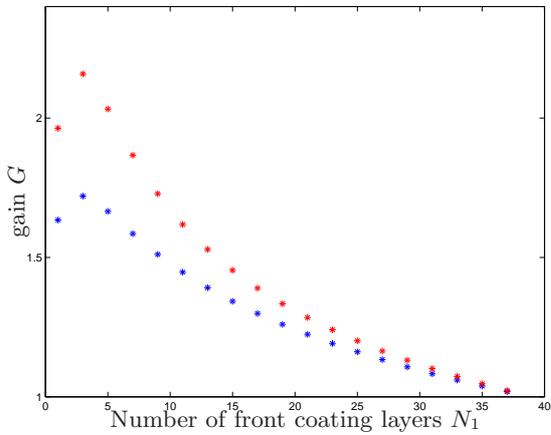}
\end{center}
\caption{Gain $G$ (see \eqref{eq_gain}) in ET (blue stars) and aLIGO (red stars) as a function of the number of front coating layers $N_1$. The number of rear coating layers $N_2$ results from the total number of layers $N=38$ minus the number of front layers $N_1$. Parameters used for this calculation are presented in Tables \ref{paramGeom} and \ref{paramPhys}. For ET a maximum gain of $G=1.72$ appears for $N_1=3$ front coating layers. For aLIGO the maximum gain of $G=2.16$ is also obtained for $N_1=3$.}\label{gain}
\end{figure}

First of all, we analyze the spectral density of the displacement noise $S_\text{total}^\text{(CM)}$ for a Khalili etalon (KE) as a function of the number of front layers $N_1$. 
This noise analysis considers the sum of the noise sources listed in Sec.~\ref{N1Opt}: Brownian, TE and TR noises which are divided into a coating and a substrate contribution each.
The KE total thermal noise is then compared to the results for a conventional mirror (CM) $S_\text{total}^\text{(CM)}$ using a {\em gain} parameter $G$ which is defined as:
\begin{equation}\label{eq_gain}
G= \left.\sqrt{\frac{S_\text{total}^\text{(CM)}(\omega)}{S_\text{total}^\text{(KE)}(\omega)}
	}\right|_{\omega=2\pi\, 100\, \text{s}^{-1}}.
\end{equation}
This gain $G$ has to be maximized in order to enhance the detector sensitivity. 
In Fig.~\ref{gain} we plot the gain $G$ as a function of the number of front coating layers $N_1$ for both, ET and aLIGO. Please recall that the number of rear coating layers $N_2=N-N_1$ is constrained by the total number of coatings $N=38$.
One can see that for both detectors, ET and aLIGO, the gain is obviously maximized for the case of $N_1=3$ front coating layers. Then the maximum gain appears to be $G=1.72$ for ET and $G=2.16$ for aLIGO.

\subsection{Absorption}\label{optAbs}
Another important parameter to be taken into account is the substrate absorption $\Psi_\text{abs}$.
It describes the portion of light energy which is absorbed in the substrate with respect to the light incident to the mirror. 
The value of $\Psi_\text{abs}$ is accessible via the light power circulating inside the substrate.
We know that this light power inside the etalon will be a factor of
\begin{equation*}
\frac{1-R_1^2}{(1+R_1R_2)^2}
\end{equation*}
lower than the light power incident to the mirror.
Finally, the value $\Psi_\text{abs}$ may be evaluated using the absorption coefficient of the substrate material $\eta$ (in ppm per cm) and the substrate thickness $h$ (in cm): 
\begin{equation}\label{SubAbs}
\Psi_\text{abs}=2\eta h\frac{1-R_1^2}{(1+R_1R_2)^2}.
\end{equation}
Note the factor of 2 in front of the substrate thickness, which occurs as the light passes the substrate twice: once forward and once backward.
Also keep in mind that the reflectivities $R_1$ and $R_2$ are functions of the number of front and rear coating layers --- see formulas (\ref{R1R2}).

\begin{figure}[htbp]
\begin{center}
\psfrag{Einstein Telescope and Advanced LIGO KE losses}[cc][cc]{\small }
\psfrag{Substrate absorbtion}[cc][cc]{\small substrate absorption $\Psi_\text{abs}$, ppm}
\psfrag{Number of front coating layers}[cc][cc]{\small Number of front coating layers $N_1$}
\includegraphics[width=0.4\textwidth]{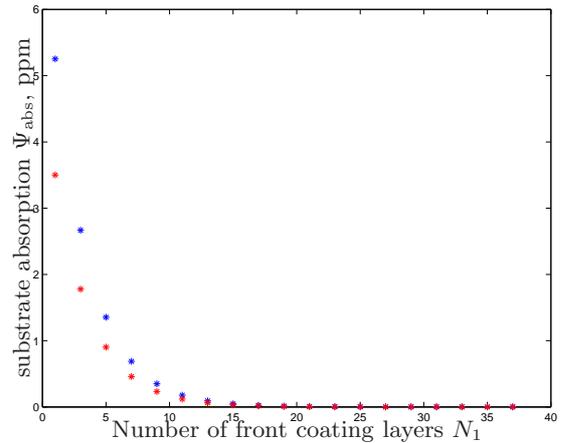}
\end{center}
\caption{Substrate absorption $\Psi_\text{abs}$ in ET (blue stars) and aLIGO (red stars) as a function of the number of front coating layers $N_1$. The number of rear coating layers $N_2$ results from the total number of layers $N=38$ minus the number of front layers $N_1$. Parameters used for this calculation are presented in Tables \ref{paramGeom} and \ref{paramPhys}.}\label{losses}
\end{figure}

As Fig.~\ref{losses} illustrates $\Psi_\text{abs}$ decreases exponentially with an increasing number of front coating layers. 
For the optimum number of front coating layers $N_1=3$ the substrate absorptions in ET and aLIGO equal $2.67$~ppm and $1.78$~ppm, respectively. 
It seems reasonable to assume that a loss coefficient of $1$~ppm is admissible. 
In this case we have to choose $N_1=5$. 
Indeed, using formula (\ref{R1R2}) with $N_1=5$ and coating parameters listed in Tables \ref{paramGeom} and \ref{paramPhys} we obtain:
\begin{align}
 \Psi_\text{abs}^\text{(AL)}&\simeq 0.90\ \text{ppm},\quad
 \Psi_\text{abs}^\text{(ET)}\simeq 1.35\ \text{ppm}.
\end{align}
It means that for aLIGO (circulating power $W_0=0.8$~MW) the absorbed power is about $W_\text{abs}^\text{(AL)}\simeq 0.72$~W, as for ET ($W_0=3$~MW) --- $W_\text{abs}^\text{(ET)}\simeq 4.1$~W.

Consequently, everywhere below in this article we assume the number of front coating layers to be $N_1=5$. 
This choice allows the gain to be $G=1.67$ for ET and $G=2.03$ for aLIGO. 

\begin{table}[htbp]
\begin{center}
\caption{Parameters used for the numerical calculation. $w$ is the radius of the laser beam (intensity decreases with distance $r$ from the center as $\sim e^{-2r^2/w^2}$), $R$ and $h$ are radius and thickness of the cylindrical mirror, $W_0$ is the power circulating in the arm cavities of the interferometer.}\label{paramGeom}
\begin{tabular}{|c|c|c|}
\hline
~Parameter~& ~Einstein Telescope & ~Advanced LIGO \rule{0mm}{5mm}\\[0.5mm]
\hline
$w$, m & $0.12$ & $0.06$\cite{2004_TDInhomog_Fejer} \\
$R$, m & $0.31$ & $0.17$ \\
$h$, m & $0.30$ & $0.20$ \\
$W_0$, MW & $3$ & $0.8$\cite{2007_aLIGO_param}\\
$N^\text{(CM)}$ & $39+$cap & $39+$cap \\
$N^\text{(KE)}$ & $38+2$ caps & $38+2$ caps \\
\hline
\end{tabular}
\end{center}
\end{table}

\begin{table}[htbp]
\begin{center}
\caption{Parameters used for the numerical calculations. $T$ is the temperature, $\Lambda$ is the optical wavelength in vacuum, $\eta$ denotes the optical losses per unit length in the substrate and $n$ is the refractive index. The remaining parameters are explained in Sec.~\ref{optBN}-\ref{optTRN}.
}\label{paramPhys}
\begin{tabular}{|c|c|c|c|}
\hline
~Parameter~& ~substrate & ~\tao\ layer & ~\sio\ layer \rule{0mm}{5mm}\\[0.5mm]
\hline
$T$, K &\multicolumn{3}{c}{300}\cite{2004_TDInhomog_Fejer}~\vline\\
$\Lambda$, m &\multicolumn{3}{c}{$1.064\times10^{-6}$}\cite{2004_TDInhomog_Fejer}~\vline\\
\hline
$\eta$, ppm/m &~ $25$ &~ - &~ -\\
$n$\cite{2004_TDInhomog_Fejer} &~ 1.45 &~ 2.035 &~ 1.45\\
$\beta$, 1/K \cite{2009_TermoOpt_Evans,2004_TDInhomog_Fejer} 
&~ $8\times10^{-6}$ &~ $1.4\times10^{-5}$ &~ $8\times10^{-6}$\\
$\alpha$, 1/K\cite{2004_TDInhomog_Fejer} &~ $5.1\times10^{-7}$ &~ $3.6\times10^{-6}$ &~ $5.1\times10^{-7}$\\
$\rho$, kg/m$^3$\cite{2004_TDInhomog_Fejer} &~ $2202$ &~ $6850$ &~ $2202$\\
$Y$, Pa &~ $72\times10^{9}$\cite{2004_TDInhomog_Fejer} &~ $140\times10^{9}$\cite{1994_FilmsProps_Martin} &~ $72\times10^{9}$\cite{2004_TDInhomog_Fejer}\\
$\nu$ &~ $0.17$\cite{2004_TDInhomog_Fejer} &~ $0.23$\cite{1994_FilmsProps_Martin} &~ $0.17$\cite{2004_TDInhomog_Fejer}\\
$\kappa$, W/K m\cite{2004_TDInhomog_Fejer} &~ $1.38$ &~ $33$
 &~ $1.38$\\
$C$, J/K kg\cite{2009_TermoOpt_Evans} &~ $746$ &~ $306$ &~ $746$\\
$\phi$ &~ $4\times10^{-10}$\cite{2006_LossAngleSi_Penn} &~ $2\times10^{-4}$\cite{2007_LossTa_Harry} &~ $4\times10^{-5}$\cite{2003_LossTaSi_Penn}\\
\hline
\end{tabular}
\end{center}
\end{table}

\subsection{Thermal noise of the Khalili etalon}\label{TNinKE}

\begin{figure}[htbp]
\begin{center}
\includegraphics[width=0.5\textwidth]{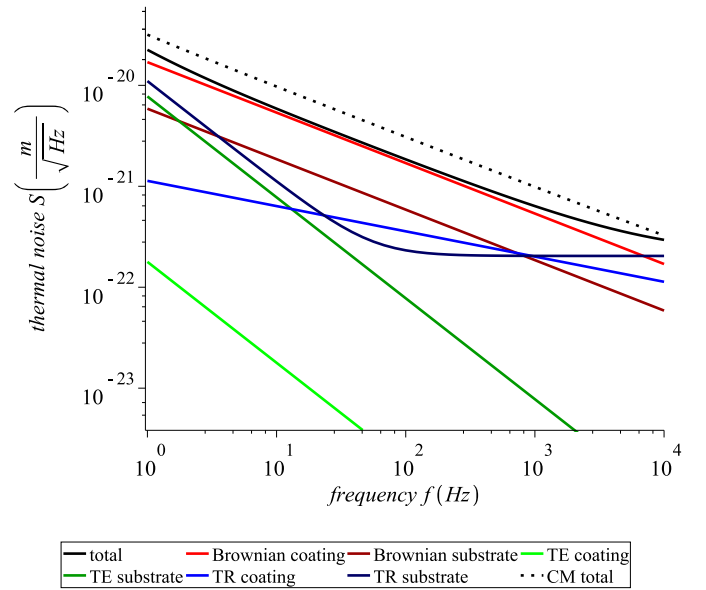}
\end{center}
\caption{ET noise spectral densities for KE with a front coating of $N_1=5$ layers (plus a cap) and a rear coating of $N_2=33$ layers (plus a cap). Parameters used for this calculation are presented in Table~\ref{paramGeom} and Table~\ref{paramPhys}.}\label{ETsense}
\end{figure}

\begin{figure}[htbp]
\begin{center}
\includegraphics[width=0.5\textwidth]{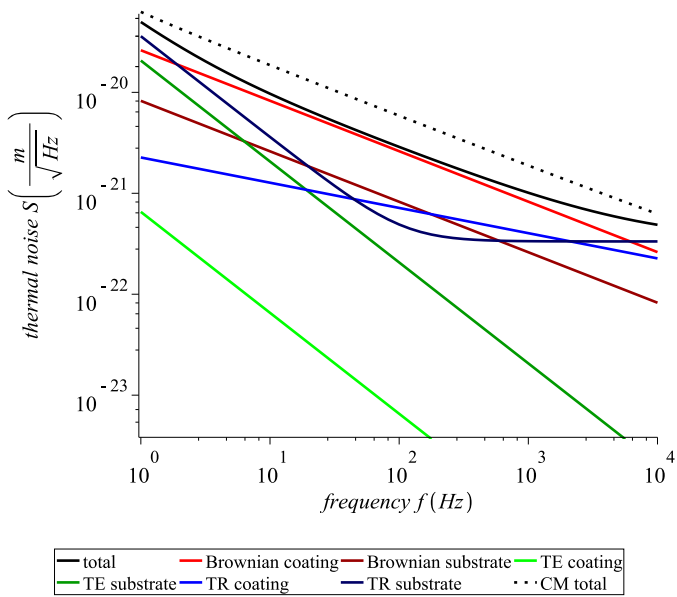}
\end{center}
\caption{aLIGO noises spectral densities for KE with a front coating of $N_1=5$ layers (plus a cap) and a rear coating of $N_2=33$ layers (plus a cap). Parameters used for this calculation are presented in Table~\ref{paramGeom} and Table~\ref{paramPhys}.}\label{ALsense}
\end{figure}

In Fig.~\ref{ETsense} and~\ref{ALsense} we present the thermal noise spectrum of a KE including different noise sources for ET and aLIGO, respectively. 
For numerical estimates we use the parameter data listed in Tables~\ref{paramGeom} and~\ref{paramPhys}. 
One clearly realizes that Brownian thermal noise dominates the mirror thermal noise at almost all the frequency range from $1$~Hz to $10$~kHz.

Seperately, we also present the numerical results for all noise sources at a single frequency of $f=100$~Hz in Table~\ref{noisesETAL}. 
We choose this frequency as a round number located in the frequency range with the highest sensitivity. 
This frequency value has already been used previously in this article in Sec.~\ref{optRes} for numerical estimates.

\begin{table}[htbp]
\begin{center}
\caption{Numerical values for all noise sources (Brownian, TE and TR noise of mirror substrate and coating) and total mirror thermal noise at $100$~Hz for ET and aLIGO. Parameters used for this calculation are presented in Tables~\ref{paramGeom} and~\ref{paramPhys}. The front coating of the KE consists of $N_1=5$ layers (plus a cap) while the rear coating consists of $N_2=33$ layers (plus a cap).
}\label{noisesETAL}
\begin{tabular}{|l|c|c|}
\hline
~noise spectral density~& ~Einstein Telescope & ~Advanced LIGO \rule{0mm}{5mm}\\[0.5mm]
\hline
Khalili Etalon (KE):&&\\
coating Brown., $\frac{\text{m}}{\sqrt{\text{Hz}}}$ &$1.70\times10^{-21}$ &$2.63\times10^{-21}$ \\
substrate Brown., $\frac{\text{m}}{\sqrt{\text{Hz}}}$ &$6.06\times10^{-22}$ &$8.55\times10^{-22}$ \\
coating TR, $\frac{\text{m}}{\sqrt{\text{Hz}}}$ &$3.58\times10^{-22}$ &$7.16\times10^{-22}$ \\
substrate TR, $\frac{\text{m}}{\sqrt{\text{Hz}}}$ &$1.60\times10^{-22}$ &$3.38\times10^{-22}$ \\
substrate TE, $\frac{\text{m}}{\sqrt{\text{Hz}}}$ &$7.98\times10^{-23}$ &$2.14\times10^{-22}$ \\
coating TE, $\frac{\text{m}}{\sqrt{\text{Hz}}}$ &$1.79\times10^{-24}$ &$6.63\times10^{-24}$ \\
\hline
KE total, $\frac{\text{m}}{\sqrt{\text{Hz}}}$ &$1.85\times10^{-21}$ &$2.89\times10^{-21}$ \\
\hline
Conv. Mirror (CM):&&\\
coating Brown., $\frac{\text{m}}{\sqrt{\text{Hz}}}$ &$2.99\times10^{-21}$ &$5.75\times10^{-21}$ \\
substrate Brown., $\frac{\text{m}}{\sqrt{\text{Hz}}}$ &$6.53\times10^{-22}$ &$9.32\times10^{-22}$ \\
coating TR, $\frac{\text{m}}{\sqrt{\text{Hz}}}$ &$3.58\times10^{-22}$ &$7.16\times10^{-22}$ \\
substrate TE, $\frac{\text{m}}{\sqrt{\text{Hz}}}$ &$8.51\times10^{-23}$ &$2.31\times10^{-22}$ \\
coating TE, $\frac{\text{m}}{\sqrt{\text{Hz}}}$ &$4.52\times10^{-24}$ &$1.78\times10^{-23}$ \\
\hline
CM total, $\frac{\text{m}}{\sqrt{\text{Hz}}}$ &$3.09\times10^{-21}$ & $5.88\times10^{-21}$ \\
\hline
\end{tabular}
\end{center}
\end{table}

Brownian noise is the main object of our investigations as it dominates the sensitivies of both detectors (ET and aLIGO) almost in the whole detection band. 
It can be calculated accurately using the model developed in~\cite{2010_KE_Somiya}. 
Shortly, a calculation method is presented in Sec.~\ref{optBN}. 
A further inspection of the spectral noise sensitivity plots reveals substrate Brownian noise to be the second important noise process. Thus, Brownian noise dominates over TE and TR noise absolutely and is the main factor limiting both interferometers sensitivities in the frequency domain near $100$~Hz. 

Taking the same noise sources into account as for our KE investigation (see Sec.~\ref{optBN}-\ref{optTRN}) we have calculated the numerical values for the spectral noise density of a corresponding CM.
For clarity we do not present the single contributions but only the total mirror thermal noise in this section. 
Using thermal noise values for KE and CM ($S_\text{tot}^\text{(KE)}$ and $S_\text{tot}^\text{(CM)}$) and the definition of the gain parameter $G$ \eqref{eq_gain} we compare KE and CM to estimate the benefit of using a KE. 
For the spectral densities calculated in Table~\ref{noisesETAL} we arrive at a gain of 
\begin{align}
\label{Getav}
G_\text{ET} & =1.67,\ \text{for ET parameters},\\
G_\text{AL} & =2.03,\ \text{for aLIGO parameters}.
\end{align} 
The sligtly larger gain value for aLIGO parameters may be qualitatively explained by the mirror geometry.
So the ET mirror is more sensitive to membrane deformations than the aLIGO mirror. 
This property allows to 'transfer' Brownian fluctuations from the rear coating to the front surface more effectively. 
Indeed, the geometrical factor $g=2R/h$ (the fraction of diameter to thickness of mirror) for ET is larger than for aLIGO: 
\begin{align}
\label{g}
 g_\text{AL}\simeq 1.7,\quad g_\text{ET}\simeq 2.1 .
\end{align}

\subsubsection{Semi-qualitative consideration}\label{SimCon}
In this subsection we would like to present a way to simply estimate the gain. 
For an order of magnitude estimate we may approximate the total thermal noise by Brownian coating noise that prevails at all frequency ranges as we have seen. 
Moreover, the thickness of \sio\ layers and \tao\ layers differs only about $40$\,\% while the \tao\ loss angle is $5$ times higher than the loss angle of \sio. 
Therefore, we may very roughly approximate the total thermal noise with the sum of \tao\ coating layers Brownian noise (recall that all of them are uncorrelated). 
In a first approximation one could assume that the front coating is responsible for the main contribution to the noise level. Thus, the total thermal noise level should be proportional to the number of front coating \tao\ layers only. 
For a CM this number is $Q_\text{CM}=20$ and for a KE -- $Q_\text{KE}^\text{front}=3$. 
The ratio of the $Q$ values should represent the gain of a KE:
\begin{equation}\label{GappKC}
G_\text{naive}=\sqrt{\frac{Q_\text{CM}}{Q_\text{KE}^\text{front}}}=\sqrt{\frac{20}{3}}=2.58.
\end{equation}
The estimated gain is larger compared to the accurate results (\ref{Getav}). 
It may be explained by the fact that we do not account for elastic coupling (through substrate) between rear coating layers motion and front coating layers motion, i.e. the displacement of the front coating due to a deformation of the rear coating layers.
One could say, the rear coating layers motion is {\em 'transferred'} to the front coating through the substrate.
This coupling is moderated by the elastic properties of the latter.

We introduce a transfer ratio $p$ to account for this elastic coupling.  
The variable $p$ ranges from $0$ for a Khalili cavity (KC) to $1$ for a CM or 'zero'-thickness substrate. 
We can calculate $p$ using the simple model of a cylindrical mirror whose front and rear surface are covered by equal layers (same thickness and same elastic parameters).
Let us apply a single force at the front surface and keep the rear surface free of forces. 
One can calculate the elastic energies in the front layer $U_\text{front}$ and in the rear layer $U_\text{rear}$.
Obviously, the transfer ratio $p$ may be calculated as the ratio of both energies. 
This estimate gives:
\begin{align}
p &\equiv \frac{U_\text{rear}}{U_\text{front}},\quad p_\text{ET}\simeq 0.23,
 	\quad p_\text{AL}\simeq 0.086 
\end{align}
The ratio $p_\text{AL}$ for aLIGO is smaller than for ET. 
Again this behaviour can be explained by the different geometry factors $g$ (see estimates (\ref{g})).

Now instead of Eq.~(\ref{GappKC}) we can state a more accurate formula for the gain estimate taking into account the elastic coupling $p$ of the $Q_\text{KE}^\text{rear}=17$ rear coating layers:
\begin{subequations}\label{GappKE}
\begin{align}
G_\text{app}&=\sqrt{\frac{Q_\text{CM}}{Q_\text{KE}^\text{front}+p\times Q_\text{KE}^\text{rear}}},\\
G_\text{app}^\text{(ET)}&\simeq 1.71,\quad
	G_\text{app}^\text{(AL)} \simeq 2.12.
\end{align}
\end{subequations}
We see that the approximated gain values coincide with the accurate values (\ref{Getav}) within an accuracy of about $5$\,\%. 
Real gains in ET and aLIGO are lower than the expected approximated values \eqref{GappKE} because of other noise sources that were omitted here (Brownian substrate and Brownian coating of the \sio\ layers, substrate and coating TE and TR noise).

Note that the elastic coupling does not take place in a KC where both coatings are mechanically separated by vacuum. 
Consequently for both detectors, ET and aLIGO, the usage of a KC instead of a CM is expected to show a gain value of $G\approx2.6$ (as \eqref{GappKC}).

\subsection{Potential sensitivity improvements for future GW interferometers}
In this section we quantitatively analyse the overall sensitivity improvement potentially achievable by replacing the conventional end mirrors by KE in aLIGO and ET.

\begin{figure}[htbp]
\begin{center}
\includegraphics[width=0.5\textwidth]{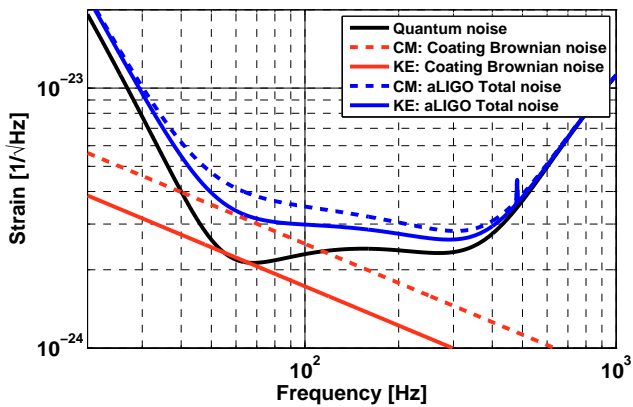}
\end{center}
\caption{Potential overall sensitivity improvement of aLIGO for the use of KE,
compared to conventional end mirrors: The binary neutron star inspiral range increases by 15\,\%
from 196.2 to 225.0\,Mpc. This corresponds to a potential increase in the detected event rate 
for binary neutron star inspirals of slightly above 50\,\%. }\label{ALIGO_comparison}
\end{figure}

\begin{figure}[htbp]
\begin{center}
\includegraphics[width=0.5\textwidth]{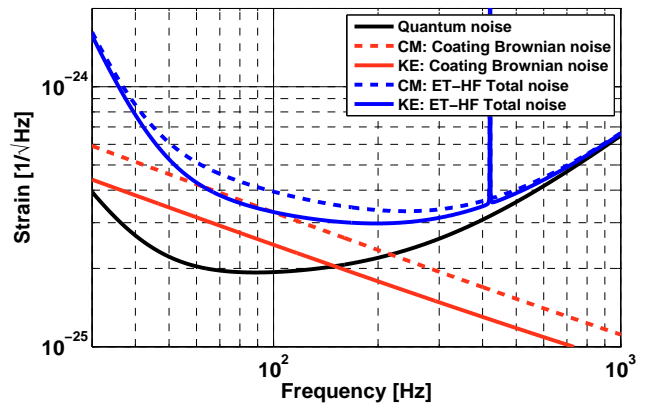}
\end{center}
\caption{Potential overall sensitivity improvement of the ET high frequency detector 
(as described in \cite{2010_Hild_xyl}) for the use of KEs,
compared to conventional end mirrors: The binary neutron star inspiral range increases by 15\,\%, 
corresponding to a potential increase in the detected event rate 
for binary neutron star inspirals of 50\,\%. }\label{ET_comparison}
\end{figure}

Figure \ref{ALIGO_comparison} shows the potential sensitivity improvement of aLIGO 
for the use of KC as end mirrors. The sensitivity curves have been created using
 the GWINC software \cite{website_GWINC} and for a signal recycling configuration
 that is optimised for the detection of binary neutron star inspirals (see configuration 2 in \cite{2008_aLIGO_design}).
 Only the two main noise contributions are shown: quantum noise (black trace) and coating Brownian
 noise (sum of all test masses) (red trace), as well as the total noise (blue traces).
 Please note that all other relevant noise sources have been included in the calculations of the total noise traces, but 
 have been omitted from the plot for clarity. The dashed lines indicate the strain levels for the standard 
 aLIGO design, while the solid lines show the potentially reduced noise levels originating from the
 application of KE as end test masses, as described in this article. The main difference between these two scenarios
 originates from the reduction of coating Brownian noise by a factor 2.18, as described by the values
  in the right hand column of Table \ref{noisesETAL}. Please note that thermal noise contributions
  from the input mirrors stay identical for the two scenarios. 
The corresponding increase in the binary inspiral range (1.4 solar masses, SNR of 8, averaged sky location)
is about 15\,\% and therefore yields an relative increase of the binary neutron star inspiral event rate of 
about 50\,\%.

Figure \ref{ET_comparison} shows the sensitivity improvement of a potential ET high frequency detector as described in \cite{2010_Hild_xyl} for the replacement of the 
conventional end mirrors by KEs. Following the values given in Table \ref{noisesETAL} 
we considered a flat coating Brownian noise
reduction factor of 1.76 for the end test masses, while again we assumed the thermal noise
of the input test masses to stay constant. This yields an overall reduction of the total thermal 
noise of all test masses of about 25\,\% and an increase in the observatory sensitivity of up to 
20\,\% in the most sensitive frequency band between 50 and 400\,Hz. We find an increase in 
the binary neutron star inspiral range of 15\,\% from 1593 to 1833\,Mpc. This corresponds to an increase 
in the binary neutron star inspiral event rate of about 50\,\%.

\section{Technical feasability of Khalili etalons for future GW observatories }\label{ThermLens}

\subsection{Required Hardware}

Figure \ref{fig:schematics} shows the simplified schematics of an aLIGO
or ET interferometer with different end mirror configurations. 
Replacing the conventional end mirrors by KCs would have a significant
impact on the required hardware. Instead of a single end 
mirror suspended from a single seismic isolation system per end mirror, in the case 
of the KC two mirrors with two full seismic isolation systems are required
at the end of each arm cavity.

This means that with KCs there are six (2x IM, 2x IEM, 2x EEM) instead 
of four optical elements (2x IM, 2 EM), which require the maximal seismic isolation. 
The concept of the KE allows us to still significantly reduce the thermal noise 
contribution of the end mirrors, while being compatible with the already available seismic
isolation systems. Therefore, it is in principle possible to upgrade a 2nd or 3rd generation
gravitational wave detector by replacing conventional end mirrors by KEs without altering or extending the vacuum systems and seimsic isolation systems.

\subsection{Interferometric Sensing and Control}

In addition to the reduced hardware requirements, the main advantage of the KEs with respect to KCs is the potential simplification of several aspects
related to the interferometric sensing and control. Upgrading aLIGO or an 
ET interferometer from its standard configuration to employ KCs increases
the length degrees of freedom of the main interferometer from five (DARM (differential arm length), MICH (Michelson cavity length), SRCL (signal recycling cavity length), CARM (common arm length), PRCL (power recycling cavity arm length)) to a total of seven.

 It is worth mentioning that the additional two degrees of freedom
actually have a very strong coupling to the differential arm length channel of the interferometer.
For the example of the coating distribution discussed in this article, the length of the KC needs to be stabilized with an accuracy of only a factor 10 less than what is required for 
 the main arm cavities. That means the length of the KC needs to be orders of magnitude 
 more stable than for example the differential arm length of the central Michelson interferometer.
 
\begin{figure*}[htbp]
\begin{center}
\includegraphics[width=1\textwidth]{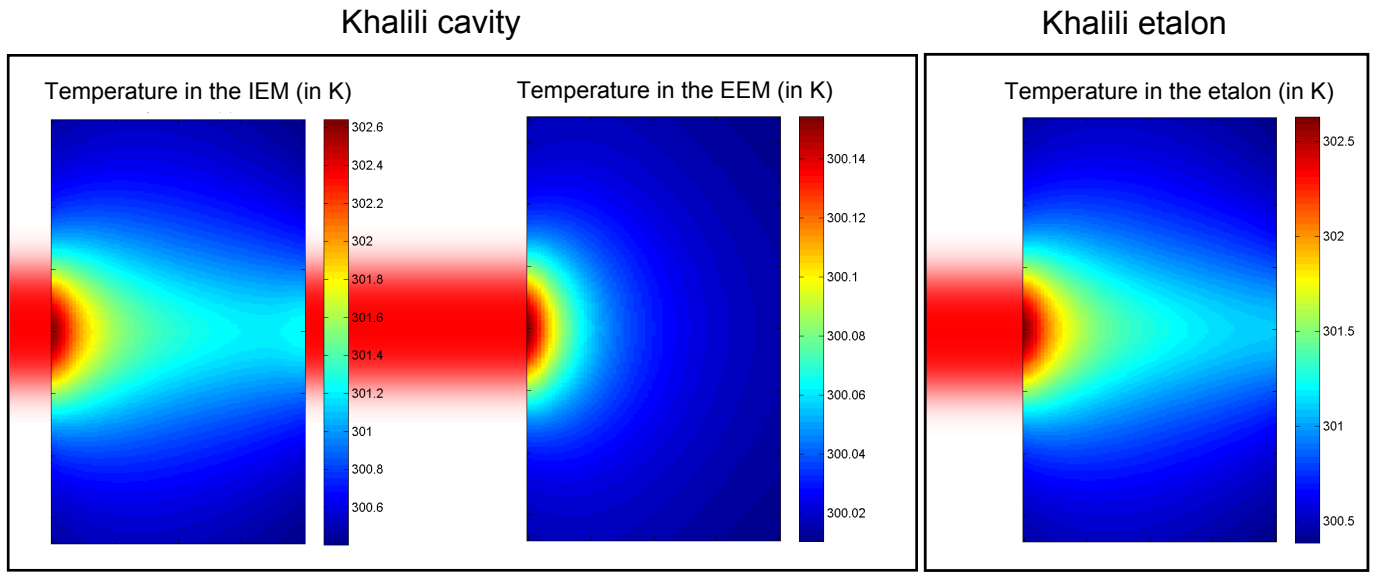}
\end{center}
\caption{The temperature distribution in the mirrors of a KC (left) and a KE (right) for aLIGO parameters (with N$_1$=5) as calculated via FEM. Please note the different color scales for the IEM and the EEM in the KC.}\label{fig:FEM}
\end{figure*}

 In order to achieve this demanding stability of the KC one has to make use of highly
 dedicated readout and control schemes. Special care needs to be taken to avoid introducing 
 potential control noise at the low frequency end, which could potentially spoil the overall sensitivity
 of the gravitational wave detector.
 
 Substituting the KC, consisting of two individual mirrors potentially encountering independent
 driven motion (for example seismic), by the proposed KE would ensure that both relevant 
 mirror surfaces would be rigidly coupled via the etalon substrate. Therefore, the length of the KE would be much 
 less susceptible to seismic disturbances or gravity gradient noise, as compared to the length of KC.
 Also in terms of potential control noise the KE is advantageous over the KC. In
 case of the KC the mirror positions would have to be controlled by means of coil magnet actuators
 or electro-static actuators, which can potentially introduce feedback noise at frequencies within the 
 detection band of the gravitational wave detector. In contrast the length of the KE can be locked
 by controlling the etalon's substrate temperature (using the temperature dependency of the index of refraction). 
Since the etalon substrate acts as a thermal low pass, the etalon length will be extremely constant for 
all frequencies within the detection band of the interferometer.

However, not only the length sensing and control is highly demanding in case of a KC, but 
also the alignment sensing and control. Again the key point here is to find a high signal to noise
error-signal and then applying low noise feedback systems to keep the mirrors of the KC 
aligned in pitch and yaw. As the KC would be rather short compared to the main arm cavities,
the KCs would unfortunately feature a high mode degeneracy, i.e. it would not only be resonant
for the desired TEM$_{00}$ mode, but also for higher order modes, which would further increase the alignment
requirements. Using a KE would potentially allow us to transfer the alignment control problem 
from the detector operation to the manufacturing process of the etalon. If it would be possible to manufacture
an KE with sufficiently parallel front and back surface, we would not need to actively control the 
relative alignment of the KE surfaces during operation. The two parameters that would be most relevant
are the relative curvature mismatch of the etalon front and rear surfaces as well as the parallelism of the two
surfaces. As we have shown in \cite{2009_Hild_Virgo_etalon} the curvature mismatch is the dominating factor
for the etalon's performance.

\begin{figure*}[htbp]
\begin{center}
\includegraphics[width=1\textwidth]{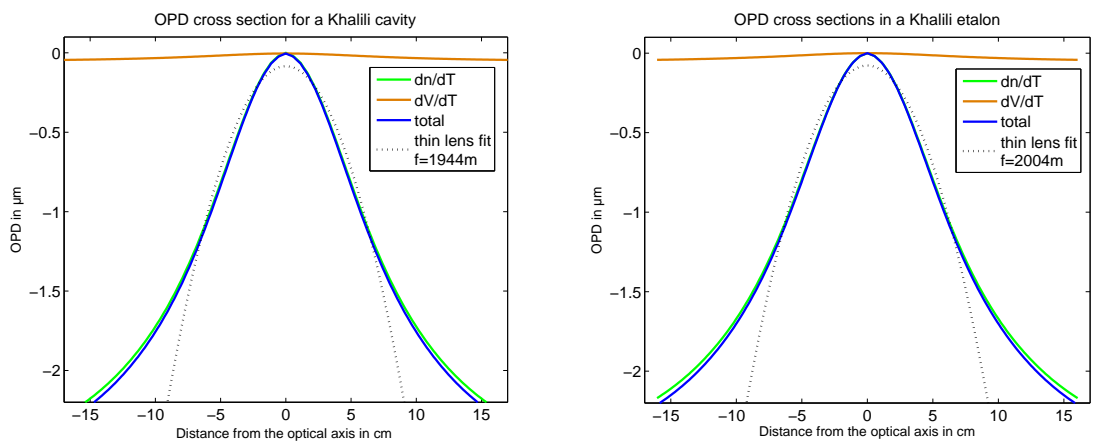}
\end{center}
\caption{The optical path delay (OPD) is computed in the FEM simulation, in order to determine the focal length of the thermal lens in the KC (left) and the KE (right) for aLIGO parameters. Shown are the total OPD, the OPD caused by the thermo-refractive effect (named $dn/dT$) and the OPD due to the expansion of the substrate ($dV/dT$). The plots show also fits of the OPD for an ideal thin lens.}\label{fig:Therm_lens}
\end{figure*}

\subsection{Thermal Lensing}

In this section we will compare the thermal lensing \cite{Walter1991} 
of the KC configuration to the one of a KE. In the case of the 
KC we have the following absorption processes: (i) IEM front coating, (ii) IEM substrate,
(iii) IEM anti-reflex coating on its rear surface and (iv) front coating of EEM. In the case of the 
proposed KE the situation is pretty similar apart from process (iii), which does not exist. 

In the following we will show by means of FEM that the actual thermal lensing induced into the 
KE is of the same order, but slightly smaller than in the case of the KC. 

 The FEM used here treats the mirror as a substrate. The coatings and the laser beam are included as heat sources. For the reflective coatings we assume an absorption of 0.5 ppm, and 1 ppm for the anti-reflective coating in the KC. The FEM assumes an emissivity $\epsilon=0.93$, an ambient temperature of 300~K and uses the parameters from Table \ref{paramGeom} and \ref{paramPhys} for the values of aLIGO. After computing the temperature and displacements of the finite elements, the optical path difference (OPD) is derived. For the OPD we included the temperature dependence of the refractive index (which is the dominant thermal lensing effect in fused silica) and the expansion, while we omitted the elasto-optic effect. We also did not include surface to surface radiation in the KC. This would make the thermal lens worse, and is therefore safe to exclude in order to make a conservative comparison between KC and KE. 
	Fig. \ref{fig:FEM} shows the temperature distribution in a KC and a KE for the aLIGO parameters with N$_1$=5, while Fig.~\ref{fig:Therm_lens} presents the corresponding OPD for a single pass due to the thermo-optic effect and expansion of the substrate, as well as a fit of an OPD that would be caused by an ideal thin lens. The fits are least square fits, weighted by the beam intensity. For the aLIGO parameters with N$_1$=5, the thermal lensing in the KC can be described by a thermal lens with a focal length of $f=1944$ m, while the OPD in the KE can be fitted by a thermal lens with a focal length of $f=2004$ m. The respective values for ET are a focal length of 1797 m for the KC and 1838 m for KEs. 
	It follows that the induced thermal lensing in the KC and KE is of similar strength.
	 
 As we have shown above, the thermal lensing for the KE is slightly weaker than for 
 the KC. 
 
 As one can see from the magnitude of the induced thermal lensing, the compensation of this effect 
will be extremely challenging in both cases. Potential ways of mitigating the thermal lensing could
include innovative approaches such as radiative cooling \cite{2009_Kamp}
or pre-shaped mirror or etalon substrates, which feature the wrong curvature, when being cold, but
develop the correct 'shape' when operated at the designed optical power. The compensation of thermal lensing has turned out to be
more challenging than anticipated in the first generation gravitational wave detectors. Only, the practical 
experience that will be collected with the advanced detectors will allow us to realistically judge the 
feasibility of KCs as well as KEs. However, the main purpose of the thermal lensing
analysis presented here was to show that the thermal lens will not be worse, but slightly better
for the proposed KE compared to a KC.

\section{Conclusion}\label{Conc}
In this article we have investigated the main thermal noise sources arising in the mirrors of the two next generation gravitational wave detectors: Advanced LIGO and Einstein Telescope. The thermal noise sources include Brownian, thermo-elastic and thermo-refractive noise of the mirror coatings and the mirror substrate, among which the Brownian coating noise is the largest. We applied our model developed in~\cite{2010_KE_Somiya} to study the idea of the Khalili etalon to decrease the coating thermal noise and to improve the sensitivity. The optimum KE configuration minimizing the total thermal noise level was found to be with $2$~\tao\ layers and $1$~\sio\ layer plus a cap in the front coating and with $18$~\tao\ layers and $17$~\sio\ layers plus a cap in the rear coating. However, since the substrate absorption in ET with such a configuration is $8.00$~W, and that in aLIGO is $1.42$~W, our choice is not to use the optimal but a slightly different coating distribution with $N_1=5$, i.e. $3$~\tao\ and $2$~\sio\ layers plus a cap on the front surface and $17$~\tao\ and $16$~\sio\ layers plus a cap on the rear surface. The absorbed power in the substrate with such a configuration is $4.1$~W for ET and $0.72$~W for aLIGO. Such an absorption is around 1~ppm which seems to be reasonable price for the thermal noise enhancement. The total noise spectral density of ET and aLIGO can be improved by the factors of $1.67$ and $2.03$, respectively, compared with the cases of conventional end mirrors . Moreover, we have checked our numerical calculations with a very simple qualitative consideration designed to make an order of magnitude estimation. This estimation shows an agreement of better than $5$ percent with the exact numerical calculations. A use of KEs instead of conventional end mirrors would improve the detection rate of the binary neutron star inspirals with the future gravitational wave observatories by about 50\,\%. 

We also discussed the feasibility of the Khalili etalon compared with that of the Khalili cavity. The KE is more advantageous in terms of the hardware requirements. We also compared the thermal lensing effects in the KE and in the KC and found that the former is slightly better for not having the anti-reflective coatings that KC contains on the rear surface of the front mirror. In fact, the thermal lensing problem in either case is quite severe and we should explore a way to compensate the lensing effect without imposing excess noise.

In this paper we assumed that the light is reflected on the outer surface of each coating without taking into account reflections from inner layers. A more accurate analysis shown in Ref.~\cite{2010_Epsilons_GV} gives a value of coating Brownian noise slightly lower (about $10$\,\%).

Thermoelastic and thermo-refractive noises originate from a thermodynamical fluctuation of the temperature and the correlation of the two noises can be non-trivial with a certain set of parameters~\cite{2009_TermoOpt_Evans,2008_TDCompenstae_Gorodetsky}. In this paper, the correlation was ignored and we treated the two noises individually, which is not a problem as one of them is much lower than the other in the case of KE (see Table~\ref{noisesETAL}). It should be noted, however, that the optimal KE configuration could be determined in such a way that thermoelastic noise and thermo-refractive noise be negatively compensated if the mechanical loss angles of the coating materials were 10 times lower than the current values.

\acknowledgments
This work has been performed with the support of the European Commission under the Framework Programme 7 (FP7) Capacities, project Einstein Telescope (ET) design study (Grant Agreement 211743). A.G. Gurkovsky and S.P. Vyatchanin were supported by LIGO team from Caltech and in part by NSF and Caltech grant PHY-0967049 and grant 08-02-00580 from Russian Foundation for Basic Research. D.Heinert and R.Nawrodt were supported by the German Science Foundation (DFG) under contract SFB Transregio 7. S.Hild was supported by the Science and Technology Facilities Council (STFC). H.Wittel was supported by the Max Planck Society.

\appendix
\section{Coefficients $\epsilon_1$ and $\epsilon_2$ calculation}\label{AppA}
Let us consider KE as a Fabry-P\'erot interferometer with two mirrors (namely two reflective coatings) with amplitude transmittances $T_1$ and $T_2$, and amplitude reflectivities $R_1=\sqrt{1-T_1^2}$ and $R_2=\sqrt{1-T_2^2}$. The mirrors are separated by a medium with a refractive index $n_s$, and the mean distance between the mirrors is $L$. Optical losses are equal to zero. The fluctuations of the coordinates of the front and rear mirrors are reprenseted by $x$ and $y$, respectively. The probe beam $A$ is incident on the front mirror (coating) and is partially reflected. We are interested now in the reflected beam $B$. The weight coefficients $\epsilon_1$ and $\epsilon_2$ represent how much the fluctuations $x$ and $y$ contribute to the reflected beam $B$, respectively.

For a short cavity we can use a quasi-static approximation -- it means that the motion of the mirrors are sufficiently slow compared with the relaxation rate of the cavity. We assume that the optical path between the mirrors is fixed to a quarter wavelength, i.e. $e^{ikn_sL}=i$.

We can consider the cavity as a generalized mirror. Obviously, the reflectivity of the generalized mirror depends on the fluctuations $x$ and $y$. However, for the reflected beam we have to include the motion of the generalized mirror, that is, the common-mode motion of $x$ and $y$. The reflected beam shall be described as:
\begin{align}
\label{B2}
 B &= e^{2ikx}\, A\, \frac{R_2\vartheta_0^2-R_1}{1-R_1R_2\vartheta_0^2},\quad
  \vartheta_0^2=-e^{2ikn_s(y-x)}
\end{align}
We have already taken into account the fact that $e^{ikn_sL}=i$ (cavity is tuned in the anti-resonance) and $\vartheta_0$ describes the variation of cavity length due to the mirror fluctuation.

The fluctuations $x$ and $y$ are small enough compared with the cavity length $L$ so that we may expand (\ref{B2}) into series over $x$ and $y$. Keeping the linear terms only, we get
\begin{align}
 B &=-A\, \frac{R_2+R_1}{1+R_1R_2}-
  2ikA \left[x\,\epsilon_1 \frac{}{}
  +y\,\epsilon_2 \right],\\
  \epsilon_1 &=\frac{R_2(1-n_s)+R_1\big[1+(1+n_s)R_1R_2+R_2^2\big]}{\big(1+R_1R_2\big)^2},\label{e1}\\
   \epsilon_2&=\frac{n_sR_2\big(1-R_1^2\big)}{\big(1+R_1R_2\big)^2},\label{e2}
\end{align}
which have been introduced in Eq.~\eqref{e1e2}.
\section{Coating reflectivities $R_1$ and $R_2$ calculations}\label{AppB}
Let us first consider a CM with $N$ altering layers of \tao\ and \sio\ with the refractive indices $n_1$ and $n_2$, respectively, and the substrate with the refractive index $n_s$.

A multilayer coating consisting of layers with refractive indices $n_i$ and lengths $l_i$ is described by the same formulas for the transmission line consisting of ports with wave resistances $1/n_i$ and the distances $l_i$ \cite{1986_Solimeno}. 
It is convenient to describe the transmission line with impedances $Z_i$ and reflectivities $R_i$. Impedance $Z_i$ of the $i$-th layer can substitute the total impedance of all the layers between this layer and the substrate, which does not affect the other layers. 
It is convenient to start the calculation from the boundary of the substrate and the $N$-th layer, and $Z_0$ will be the equivalent impedance of all the mirror and $R_0$ will be the reflectivity of the entire system. 

There is a recurrent formula for impedances and reflectivities of neighboring layers (neighboring ports of the transmission line):
\begin{gather}
Z_i=\frac{1}{n_{i+1}}\frac{1-R_{i+1}\theta_{i}^2}{1+R_{i+1}\theta_{i}^2},\label{Zi}\\
R_i=\frac{1-n_i{\cal Z}_i}{1+n_i{\cal Z}_i},\label{Ri}
\end{gather}
where $\theta_i=e^{ikn_il_i}$ is the phase shift in the $i$-th layer. Using \eqref{Zi} and \eqref{Ri} one may easily get the recursive formula
\begin{equation}\label{ZiRecurs}
Z_i=\frac{1}{n_{i+1}}\frac{n_{i+1}Z_{i+1}^{(0)}(1+\theta_{i}^2)+(1-\theta_{i}^2)}{n_{i+1} Z_{i+1}^{(0)}(1-\theta_{i}^2)+(1+\theta_{i}^2)}
\end{equation}.

The substrate is considered as an infinite half-space, so its impedance is given by $Z_N=1/n_{N+1}=1/n_s$ and hence its reflectivity is given by $R_N=(1-n_N/n_{N+1})/(1+n_N/n_{N+1})=(n_s-n_N)/(n_s+n_N)$. Then Eqs.~\eqref{ZiRecurs} and \eqref{Ri} yields each $Z_i$ and $R_i$. We are interested only in $R_0$. The thickness of each layer in the high-reflective coating is a quarter-wavelength (QWL), i.e. $\theta_i=e^{ikn_il_i}=i$. Then \eqref{ZiRecurs} becomes:
\begin{equation}\label{ZiQWL}
Z_i=\frac{1}{n_{i+1}^2}Z_{i+1}.
\end{equation}

Using \eqref{ZiQWL} one may easily get the chain (remember that $N$ is odd and refractive indices alter so that $n_1=n_3=\dots=n_N=n_1$, $n_2=n_4=\dots=n_{N-1}=n_2$):
\begin{subequations}
\begin{align}
Z_N&=\frac{1}{n_s},\\
Z_{N-1}&=\frac{n_s}{n_1^2},\\
Z_{N-2}&=\frac{n_1^2}{n_sn_2^2},\dots\\
Z_{2m}&=\frac{n_sn_2^{(N-1)-2m}}{n_1^{(N+1)-2m}},\\
Z_{2m-1}&=\frac{n_1^{N-(2m-1)}}{n_sn_2^{N-(2m-1)}},\dots\\
Z_1&=\frac{1}{n_s}\left(\frac{n_1}{n_2}\right)^{N-1},\\
Z_0&=\frac{n_s}{n_1^2}\left(\frac{n_2}{n_1}\right)^{N-1}.
\end{align}
\end{subequations}
and thus the total coating reflectivity is:
\begin{equation}\label{Rnil}
R_0=\frac{1-n_0\frac{n_s}{n_1^2}\left(\frac{n_2}{n_1}\right)^{N-1}}{1+n_0\frac{n_s}{n_1^2}\left(\frac{n_2}{n_1}\right)^{N-1}}.
\end{equation}
The cap does not change the impedance $Z_0$. The length of the cap $l_c$ is a half-wavelength (HWL) so that $e^{ikn_cl_c}=-1$. Using \eqref{ZiRecurs} one may see that the impedance of the system with the cap is exactly the same as that without the cap: $Z_{c-1}=Z_c$. At last, Eq.~\eqref{Rnil} for $R_0$ is valid with or without the cap.

Now we are interested in reflectivities of the Khalili etalon coatings (that should be used as reflectivities $R_1$ and $R_2$ in a Fabry-P\'erot interferometer used in Appendix~\ref{AppA}). For both coatings we may use the same formula \eqref{Rnil} but with different number of layers $N_1$ and $N_2=N-N_1$ and different 'border' refractive indices $n_s$ and $n_0$:

For the {\em front coating} both the substrate and the vacuum play their original roles i.e. the rear and front infinite half-spaces respectively. Thus, $n_s=n_2$, $n_0=1$ and $N=N_1$:
\begin{equation}\label{Rone}
R_1=\frac{1-\frac{n_2}{n_1^2}\left(\frac{n_2}{n_1}\right)^{N_1-1}}{1+\frac{n_2}{n_1^2}\left(\frac{n_2}{n_1}\right)^{N_1-1}}.
\end{equation}

For the {\em rear coating} the vacuum plays the role of the substrate (the rear infinite half-space) and the substrate plays the role of the vacuum (the front infinite half space). Thus, $n_s=1$, $n_0=n_2$ and $N=N_2$:
\begin{equation}\label{Rtwo}
R_2=\frac{1-\frac{n_2}{n_1^2}\left(\frac{n_2}{n_1}\right)^{N_2-1}}{1+\frac{n_2}{n_1^2}\left(\frac{n_2}{n_1}\right)^{N_2-1}}.
\end{equation}

If we use the fact that $N_2=N-N_1$ these formulas (\ref{Rone}-\ref{Rtwo}) will coincide with formulas \eqref{R1R2}.

\end{document}